\documentclass{osa-article}

\journal{osac}

\usepackage{soul}

\begin{document}

\title{Effects of a quiet point on a Kerr microresonator frequency comb}
\author{Tomohiro Tetsumoto,\authormark{1*} Jie Jiang,\authormark{2} Martin~E.~Fermann,\authormark{2} Gabriele Navickaite,\authormark{3} Michael Geiselmann,\authormark{3} and Antoine Rolland\authormark{1}}
\address{\authormark{1}IMRA America Inc., Boulder Research Labs, 1551 South Sunset St, Suite C, Longmont, Colorado 80501, USA\\
\authormark{2} IMRA America Inc., 1044 Woodridge Ave, Ann Arbor, Michigan 48105, USA \\ 
\authormark{3}LIGENTEC SA, EPFL Innovation Park L, Chemin de la Dent d’Oche 1B, Switzerland CH-1024 Ecublens, Switzerland}
\email{\authormark{*}ttetsumo@imra.com} 

\begin{abstract} 
A quiet point, an operating point of pump-resonance detuning that minimizes frequency fluctuation due to nonlinear effects inside a resonator, has been employed for phase noise reduction of a soliton Kerr microresonator frequency comb (microcomb). Naturally, it is expected that the use of the point will also improve performances of a microcomb in terms of frequency stability and faithfulness in a phase locked loop. In this study, we experimentally investigate the effect in a microcomb with a repetition frequency of 300~GHz. We obtain a lowest fractional frequency instability at a quiet point of $1.5\times 10^{-9}$ at 1 second, which is 44 times lower than free-running instability. Phase-locking of a microcomb to a stabilized fiber comb is demonstrated to evaluate performance in a feedback loop, where in-loop-limited relative fractional frequency instability between the microcomb and the fiber comb of $6.8 \times 10^{-13}$ is obtained as an indicator of the stability limitation.
\end{abstract}


\section{Introduction}
A microcomb is a new and unique type of frequency comb source that is usually pumped with a continuous wave (CW) laser coupled to a microresonator, and has an ultrahigh repetition rate typically between 10~GHz and 1~THz~\cite{kippenberg2018dissipative}. Both bright~\cite{herr2014temporal} and dark~\cite{xue2015mode} solitons can be formed depending on resonator dispersion, and low power operation is possible on-chip with integrated ultrahigh $Q$ resonators~\cite{stern2018battery}. 
There are many applications proposed to utilize its features effectively such as optical communication~\cite{marin2017microresonator}, astronomical combs~\cite{obrzud2019microphotonic} and laser-radar~\cite{suh2018soliton,riemensberger2020massively,kuse2019frequency}. Micro-~\cite{liang2015high,liu2020photonic}, millimeter-~\cite{saleh2016phase,tetsumoto2020300,wang2021towards} and terahertz-wave~\cite{tetsumoto2020300,zhang2019terahertz} generation is one of those examples, where novel ways are actively explored to reduce phase noise and frequency instability of generated waves while keeping systems compact.
An interesting approach for that is use of a quiet point. The term of the ``quiet point'' refers to a pump-resonance detuning frequency which balances nonlinear optical effects inside a resonator (i.e., Raman-induced self-frequency shift and spectral recoil from dispersive waves or mode crossings), and gives minimized phase noise of a repetition frequency of a microcomb \cite{yi2017single}. The point can be accessed with a simple configuration for the Pound–Drever–Hall (PDH) technique~\cite{drever1983laser} and significant suppression of the phase noise can be achieved. For example, phase noise reduction of $\sim 30$~dB is observed at around 4~kHz offset experimentally for a 20~GHz microcomb based on a silicon nitride (SiN) ring resonator just by $\sim 130$~MHz of detuning tuning (supplementary information of \cite{liu2020photonic}). This implies that frequency stability of a microcomb can be improved dramatically by the method as well, however, experimental evidences have not been presented.

In this study, we empirically search for a quiet condition with pump-resonance detuning and a resonator temperature as variable parameters, and characterize three properties of a soliton microcomb: phase noise, frequency instability and faithfulness in a phase locked loop. 
First, we find interesting temperature dependence of a quite condition. We observe large changes in phase noise levels and shifts of quiet point frequencies when we tune the resonator temperature.
Next, we characterize dependence of microcomb's phase noise and frequency stability on pump-resonance detuning at an optimum resonator temperature, where important dependencies are observed for the both properties.
Finally, we perform phase-locking of a microcomb to a fiber comb to evaluate its performance in a feedback loop. The results show that the suppressed noise via the use of the quiet point helps the phase-locking operation. Limitation of frequency instability in the current configuration is also discussed.
This study will deepen understanding of behaviors of a quiet point, and light up important experimental parameters that need to be considered.

\section{Results}
\subsection{Experimental setup}
\begin{figure*}[!ht]
    \centering
    \includegraphics[width=\linewidth]{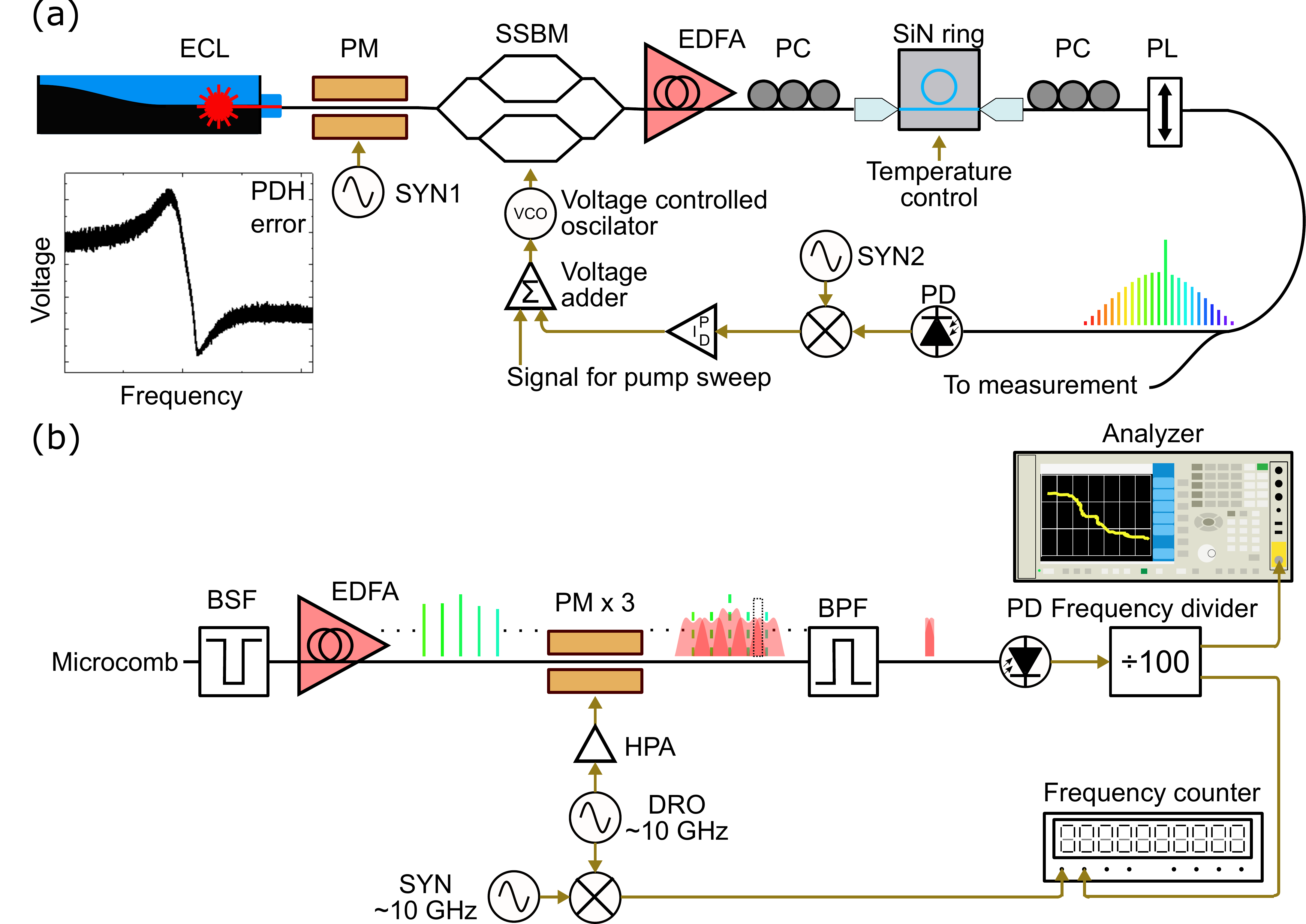}
    \caption{\footnotesize{Schematic drawing of experimental setups for (\textbf{a}) the microcomb generation and the PDH locking of pump-resonance detuning, and (\textbf{b}) phase noise and frequency stability measurement. HPA: high power amplifier. Inset at the bottom left  of (a) shows an example of the observed PDH error signal. See text for details.}}
    \label{fig1}
\end{figure*}

Figure.~\ref{fig1}(a) depicts the setup for the microcomb generation and the pump-resonance detuning control via the PDH technique. CW light from an external caivity laser diode (ECL: CTL1550, TOPTICA Photonics) is modulated with a phase modulator (PM), and its sidebands are generated. The light goes through a single sideband modulator (SSBM), is amplified with an erbium-doped fiber amplifier (EDFA) and coupled with a SiN ring resonator after polarization control with a polarization controller (PC). The photonic chip with the SiN resonator is mounted on a stage under temperature control with a Peltier device. A soliton microcomb is initiated by a fast frequency sweep of the pump light~\cite{kuse2019control}. A PC and a polarizer (PL) are installed at the output port of the microcomb to align its polarization and eliminate needs of polarization control in the following setup that consists of polarization maintaining fibers. A part of the microcomb is sent to a system for noise measurement (Fig.~\ref{fig1}(b), described later). The other part is injected into a photodiode (PD), where the modulation frequency for the PM is detected. The generated signal is mixed with a sine wave from a synthesizer (SYN2), whose frequency is identical to SYN1. A PDH error signal is obtained by synchronizing SYN2 to SYN1 and controlling their relative phase properly. An example of the error signal is shown in the inset of Fig.~\ref{fig1}(a). The error signal is applied to the frequency control channel of the SSBM through a proportional–integral–derivative (PID) controller, a voltage adder, and a voltage controlled oscillator \cite{kuse2020continuous}. By closing the PID loop, a detuning frequency between the pump and a resonance becomes same as the modulation frequency from SYN1. In this experiment, we employ detuning frequencies between 700~MHz and 1200~MHz.

Phase noise and frequency stability of the microcomb are evaluated with a system based on electro-optic (EO) combs shown in Fig.~\ref{fig1}(b) \cite{rolland2011non,tetsumoto2020300}. After the pump-power suppression with a bandstop filter (BSF) and comb-power amplification with an EDFA, EO combs of each microcomb line are generated with three cascaded PMs driven by a dielectric resonator oscillator (DRO: DRO-10010, INWAVE). Spectral overlap of two EO combs, originated from two adjacent microcomb lines, is filtered with a bandpass filter (BPF) at the center frequency of the two microcomb lines, and sent to a PD. Since the frequency of $q$~th microcomb line $f_\mathrm{q}$ is expressed by the microcomb's repetition frequency $f_\mathrm{rep}$ and carrier envelop offset (CEO) frequency $f_\mathrm{CEO}$ as $f_\mathrm{q} = qf_\mathrm{rep} + f_\mathrm{CEO}$, the detected signal gives a beat frequency of,
\begin{equation}
     f_\mathrm{beat} = (f_\mathrm{q+1} - p f_\mathrm{DRO}) - (f_\mathrm{q} + p f_\mathrm{DRO}) = f_\mathrm{rep} -  2p f_\mathrm{DRO},
    \label{equ1}
\end{equation}
where $f_\mathrm{DRO}$ is the DRO frequency, and $p$ is a minimum number of the EO comb lines required to have the spectral overlap. So, the frequency stability and the phase noise of $f_\mathrm{rep}$ can be measured from the beat signal when the frequency instability and the phase noise of the DRO are negligibly small (this condition is satisfied in measurements unless stated). The phase noise evaluation is performed with a signal analyzer (PXA n9030, Keysignt). Frequency fluctuation of the beat signal and the DRO frequency, down-converted by a SYN (HP8341A, Hewlett Packard), are measured with a frequency counter (FXE, K+K), where the SYN and the frequency counter are synchronized to a rubidium clock.

\subsection{Characterization of phase noise \& frequency instability}
\begin{figure*}[!ht]
    \centering
    \includegraphics[width=\linewidth]{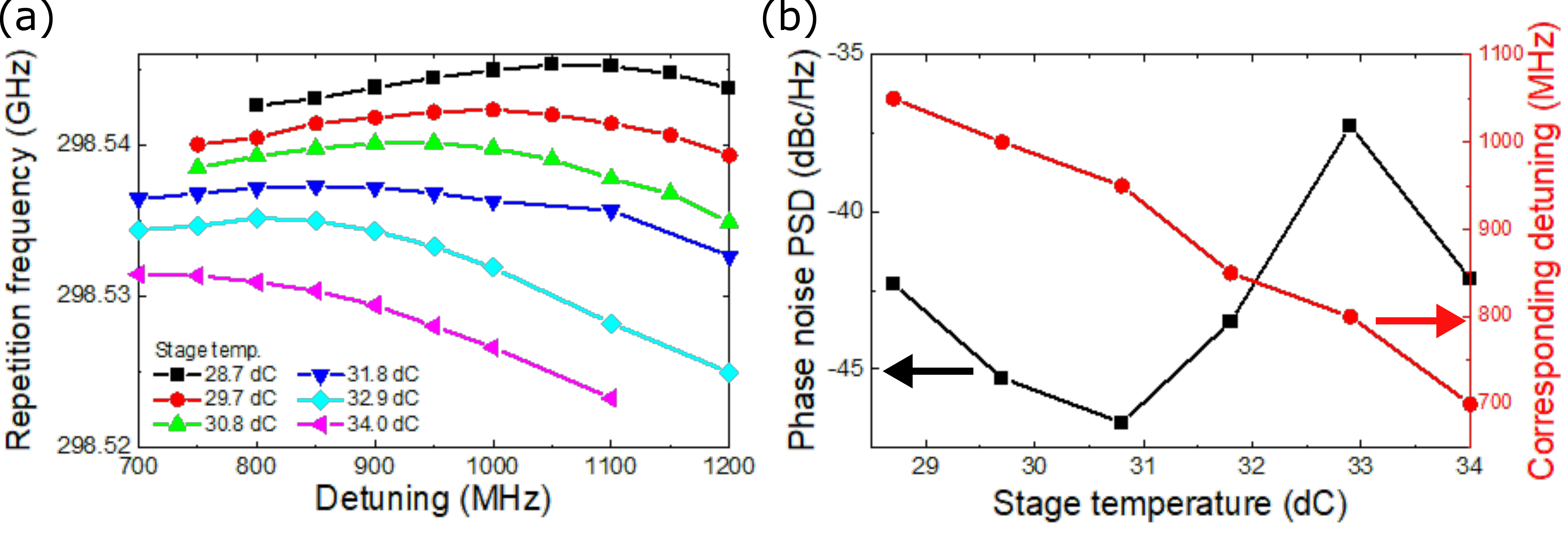}
    \caption{\footnotesize{(\textbf{a}) Pump-resonance detuning versus repetition frequency of a microcomb for various stage temperatures. (\textbf{b}) Phase noise PSD at 1~kHz offset frequency and detuning frequency of quiet points versus stage temperature.}}
    \label{fig2}
\end{figure*}

First, we investigate changes in repetition frequencies when the pump-resonance detuning is tuned. Figure~\ref{fig2}(a) present the result. The plot for the repetition frequency versus the detuning has a local maximum with $\partial f_\mathrm{rep}/\partial f_\mathrm{detuning} \sim 0$ (black curve for example), where phase noise is minimized (i.e., quiet point) as observed in the previous studies~\cite{yi2017single}. Moreover, we observe interesting shifts of detuning frequencies at local maxima when the chip-stage temperature (effectively resonator temperature) is tuned. Figure~\ref{fig2}(b) summarizes the temperature dependence of the minimized phase noise power spectral density (PSD) at 1~kHz offset and the quiet point frequencies. Difference in the minimized phase noise at each temperature is as large as 9~dB. The amount of the shift of the quiet point frequency is 350~MHz for 5~degree Celsius (dC) of the temperature change. The results suggest that to tune an operating point of the resonator temperature can be critical to minimize phase noise and have a quiet point within a reasonable operational range of the pump-resonance detuning (i.e., we cannot use too small and too large pump-resonance detuning practically). 

The observed effect should be relating to temperature dependent properties such as resonator dispersion, however, expected changes in the dispersion parameters are not so large. For instance, 0.05~\% change of a second order dispersion parameter $D_2$ (in the Taylor expansion of mode frequencies around the pump as a function of mode number such as $\omega _\mu = \omega _0 + D_1 \mu + 1/2 D_2 \mu^2 + \cdots$) is expected by 1~dC temperature change according to a calculation based on the finite element method. As the investigation of this effect is not our main scope in this study, we just add the following notes besides the report above.
Regarding the actual temperature of the resonator, we estimate that it will be tuned linearly via the stage temperature control, and the level is close to that of the stage (difference is $<0.5$ dC) in a two dimensional simulation based on the finite element method. In fact, the measured repetition frequencies get lower at a specific detuning as the temperature increases (Fig.~\ref{fig2}(a)), which is owing to increase in effective resonator length (note that we also observe the linear shift of the pump frequency, so the resonant frequency, in optical spectra). We observe nonlinear shifts of microcomb's spectral envelope center and appearances/disappearance of small spikes in optical spectra when the temperature is tuned at a specific detuning frequency. The latter will be due to mode crossings and suggests that small differences in temperature dependence of dispersion between different mode families are present, and effect the spectra. However, the strength of the spikes is not so large to move the center of mass of the spectra as dramatically as what we observed~\cite{akhmediev1995cherenkov,lucas2017detuning}. More investigation needs to be conducted to clarify the major cause of the effect.

\begin{figure*}[!ht]
    \centering
    \includegraphics[width=320pt]{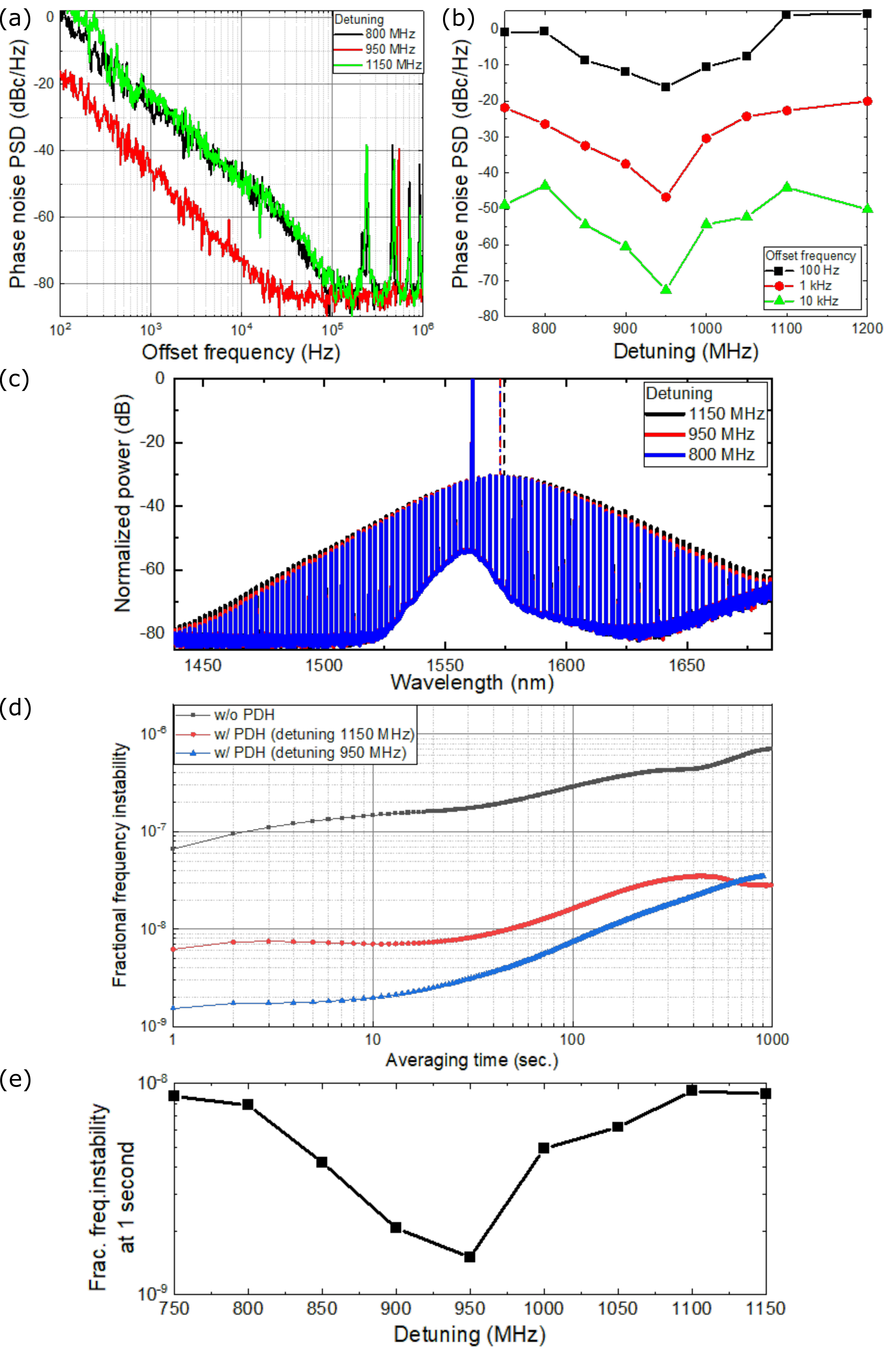}
    \caption{\footnotesize{(\textbf{a}) Phase noise PSD of the microcomb. (\textbf{b}) Phase noise PSD at offset frequencies of 100~Hz, 1~kHz and 10~kHz versus pump-resonance detuning. (\textbf{c}) Optical spectra of the microcomb. The dashed lines show the spectral envelope cneter shift from the pump. The amount of the shift is 1.41~THz, 1.43~THz and 1.60~THz for detuning of 800~MHz, 950~MHz and 1150~MHz, respectively. (\textbf{d}) Fractional frequency instability in terms of modified Allan deviation of the microcomb. (\textbf{e}) Pump-resonance detuning versus the microcomb's fractional frequency instability in terms of modified Allan deviation at 1 second averaging time.}}
    \label{fig3}
\end{figure*}

From here, we describe detailed properties of the microcomb under the fixed resonator temperature of 30.7~dC. The phase noise dependence against pump-resonance detuning is presented in Figure~\ref{fig3}(a) and (b). In Fig.~\ref{fig3}(a), three conditions of the detuning, a quiet point (950~MHz) and points with lower and higher detuning frequencies (800~MHz and 1150~MHz), are plotted. Phase noise reduction are observed over offset frequencies of around 100~Hz to 100~kHz until the plots hit the shot noise floor. Phase noise PSD at 10~kHz offset is about -73~dBc/Hz for the quiet condition. Phase noise difference between the three detuning conditions is as large as about 20~dB, 27~dB and 29~dB at 100~Hz, 1~kHz, and 10~kHz offset frequencies, respectively, as shown in Fig.~\ref{fig3}(b). Figure~\ref{fig3}(c) shows acquired optical spectra of the microcomb for the three detuning conditions. Small spectral broadening and spectral envelope's center shift from the pump are observed as the detuning frequency increases, which will attribute to the detuning dependence of the Raman-relating effect~\cite{karpov2016raman} (the spectral envelope centers are indicated by the dashed lines). Finally, we characterize instability of the repetition frequency of the microcomb. The frequency instability is lower in the PDH-locked conditions compared to that in a free-running condition as shown in Fig~\ref{fig3}(d). The obtained frequency instability at 1~second averaging time is $6.6\times 10^{-8}$, $6.2\times 10^{-9}$, and $1.5\times 10^{-9}$ for the free-running and two PDH-locked conditions with the detuning of 1150~MHz and 950~MHz, respectively. The frequency instability at the quiet point (950~MHz detuning) is the lowest among those of the detuning conditions we test as presented in Fig.~\ref{fig3}(e), and 44~times lower than that of the free-running condition. The result suggests that a quiet condition will also give minimized frequency instability as well as minimized phase noise.

\subsection{Phase-locking of a microcomb to a fiber comb}
\begin{figure*}[!ht]
    \centering
    \includegraphics[width=\linewidth]{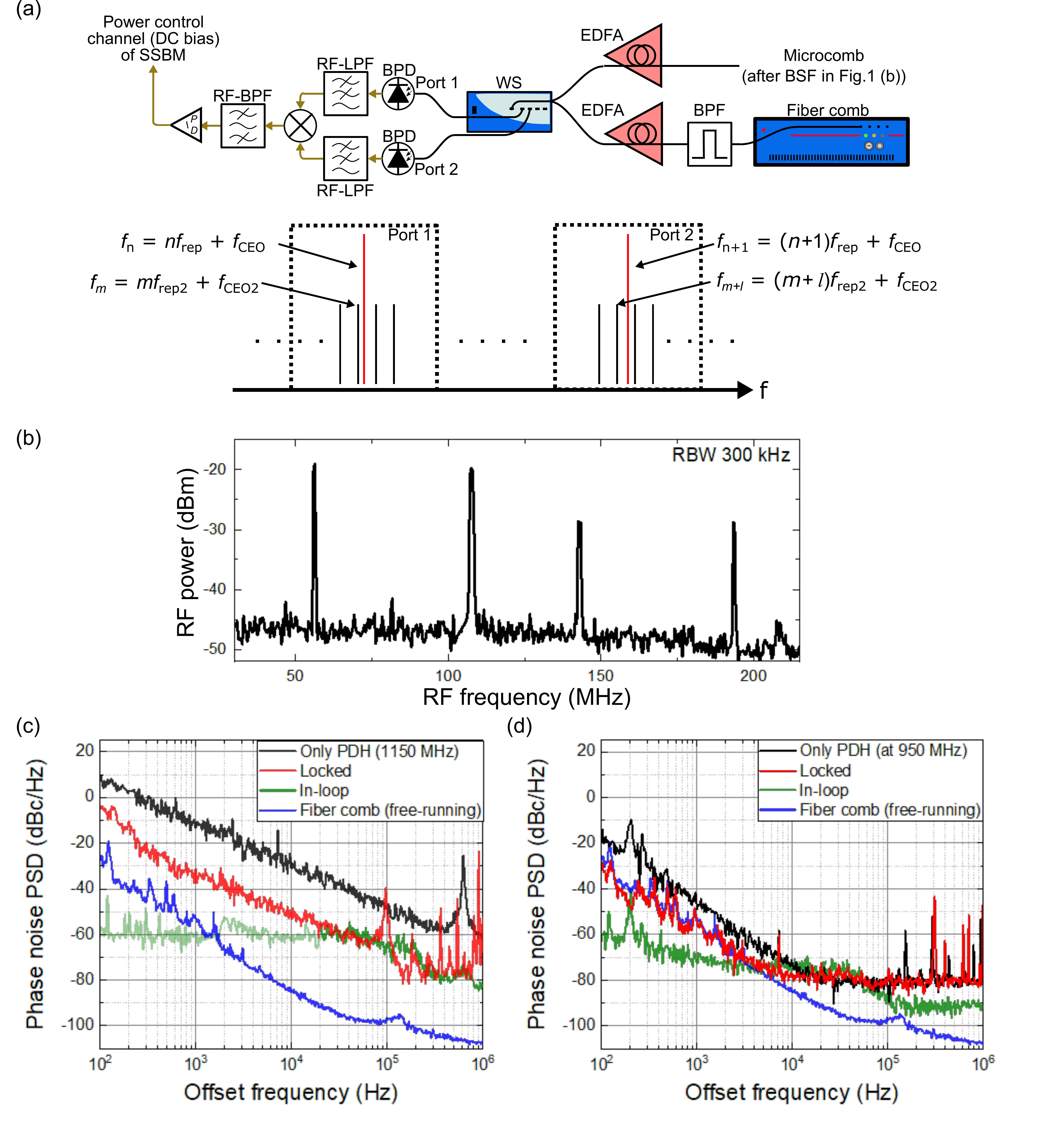}
    \caption{\footnotesize{(\textbf{a}) Schematic illustration of an experimental setup for phase-locking of a microcomb to a fiber comb (upper half), and detected optical spectrum (lower half). Red and black lines in the schematic of the optical spectrum correspond to microcomb and fiber comb lines, respectively. RF-BPF: RF bandpass filter. (\textbf{b}) RF spectrum measured after the mixer in (a). Resolution bandwidth is 300~kHz. Measured phase noise of a microcomb under (\textbf{c}) a less quiet condition with pump-resonance detuning of 1150~MHz and (\textbf{d}) a quiet condition with pump-resonance detuning of 950~MHz. The shaded region in the green line in (\textbf{c}) is not measured correctly due to the signal fluctuation.}}
    \label{fig4}
\end{figure*}
Low phase noise in a quiet condition will ease requirements in a phase-locked loop. To check the effect, we demonstrate phase-locking of a microcomb to a fiber comb. We implement a system based on detection of two beat notes between a microcomb and a fiber comb (Fig.~\ref{fig4}(a)) rather than performing the injection locking as presented in \cite{lucas2020ultralow} since the repetition frequency is rather large for it. A part of the microcomb is sampled from the point after the BSF in the Fig.~\ref{fig1}(a) and combined with comb lines from a commercial fiber comb with a repetition frequency of around 250~MHz (Ecomb-250, IMRA America). Here, only the spectral range of the fiber comb that covers 1~FSR of the microcomb is taken with a BPF to increase efficiency in the following amplification process with an EDFA. The combined light is sent to a waveshaper (WS: WaveShaper 4000A, Finisar), which is merely used as a spectral filter to send two adjacent microcomb lines with frequencies of $f_\mathrm{n} = nf_\mathrm{rep} + f_\mathrm{CEO}$ and $f_\mathrm{n+1} = (n+1)f_\mathrm{rep} + f_\mathrm{CEO}$ to two different output ports (Port~1 and Port~2). The two outputs are connected to balanced photodiodes (BPDs), respectively, where each BPD emits beat notes between each microcomb line and fiber comb lines (see the lower half of Fig.~\ref{fig4}(a)). At each port, two beat signals are generated from two fiber comb lines right next to each microcomb line at higher and lower frequnecy sides, and other signals are removed with radio-frequency low pass filters (RF-LPF). We suppose that the frequency of the fiber comb line at lower frequency side of the microcomb line $f_\mathrm{n}$ is $f_\mathrm{m}=mf_\mathrm{rep2}+f_\mathrm{CEO2}$, and that for $f_\mathrm{n+1}$ is $f_\mathrm{m+l}=(m+l)f_\mathrm{rep2}+f_\mathrm{CEO2}$, respectively, where $f_\mathrm{rep2}$ and $f_\mathrm{CEO2}$ are repetition and CEO frequencies of the fiber comb, and $m$ and $l$ are integers. Then, we obtain beat notes of $f_{1}=f_\mathrm{n} - f_\mathrm{m}$ and $f_{2}=f_\mathrm{m+1} - f_\mathrm{n}$ at Port~1 and $f_{3}=f_\mathrm{n+1} - f_\mathrm{m+l}$ and $f_{4}=f_\mathrm{m+l+1} - f_\mathrm{n+1}$ at Port~2. The signals from the two ports are mixed and four intermediate frequencies are obtained as shown in Fig.~\ref{fig4}(b). We get two intermediate signals without the CEO noise (i.e., $f_{3} -f_{1}$ and $f_{4} -f_{2}$), while the CEO noise exists in the other two (i.e., $f_{3} -f_{2}$ and $f_{4} -f_{1}$). Here, we are interested in the former case, $f_{3} -f_{1}$ for example, whose phase error is given by,
\begin{eqnarray}
     \delta _\mathrm{f_\mathrm{mixed}}   &=& \delta _\mathrm{f_\mathrm{3}} - \delta _\mathrm{f_\mathrm{1}} \nonumber\\ 
                        &=& (\delta _\mathrm{f_\mathrm{n+1}} - \delta _\mathrm{f_\mathrm{m+l})} - (\delta _\mathrm{f_\mathrm{n}} - \delta _\mathrm{f_\mathrm{m})} \nonumber\\ 
                        &=& \delta _\mathrm{f_\mathrm{rep}} - l \delta _\mathrm{f_\mathrm{rep2}},
    \label{equ2}
\end{eqnarray}
where $\delta _\mathrm{f}$ represents phase noise of a signal with frequency $f$. Only the repetition frequencies of the two combs remain in the equation, thus, we can phase-lock the microcomb to the fiber comb by constructing a phase-locked loop to cancel the noise $\delta _\mathrm{f_\mathrm{mixed}}$ as, 
\begin{equation}
    \delta _\mathrm{f_\mathrm{rep}} = l \delta _\mathrm{f_\mathrm{rep2}} +  \delta _\mathrm{f_\mathrm{loop}},
\end{equation}
where $\delta _\mathrm{f_\mathrm{loop}}$ is in-loop residual phase error. The intermediate signals without the CEO noise are distinguishable from the signals containing the CEO noise because they show relatively slim spectrum peaks.
We sample the signal at about 60~MHz in Fig.~\ref{fig4}(b) for the feedback operation and apply an error signal to a DC bias of the SSBM in Fig.~\ref{fig1}(a) to tune the input pump power~\cite{kuse2019control}.

Figure~\ref{fig4}(c) and (d) present phase noise plots in two different detuning conditions of a less quiet (1150~MHz) and the most quiet (950~MHz) points. Here, the fiber comb is under free-running operation. Phase noise levels go down in the locking conditions (red line) in the both cases, however, it does not reach the noise level of the fiber comb calibrated by the factor $l$ (blue line) in the less quiet condition (Fig.~\ref{fig4}(c)). Note that the level is even higher than that of the quiet condition without the phase-locked loop (black line in Fig.~\ref{fig4}(d)). Also, the in-loop signal was rather unstable and the measurement of its phase noise at lower offset frequencies ($<20$~kHz) was difficult (presented as a shaded green line). On the other hand, the phase noise in the quiet condition follows the calibrated fiber comb noise faithfully under the feedback control at offset frequencies of less than 4~kHz, and limited by the in-loop noise or shot noise above those offset frequencies (Fig.~\ref{fig4}(d)). These results clearly show the suppression of the phase noise via the quiet point is effective to perform a faithful operation in a phase-locked loop.

\begin{figure*}[!ht]
    \centering
    \includegraphics[width=0.8\linewidth]{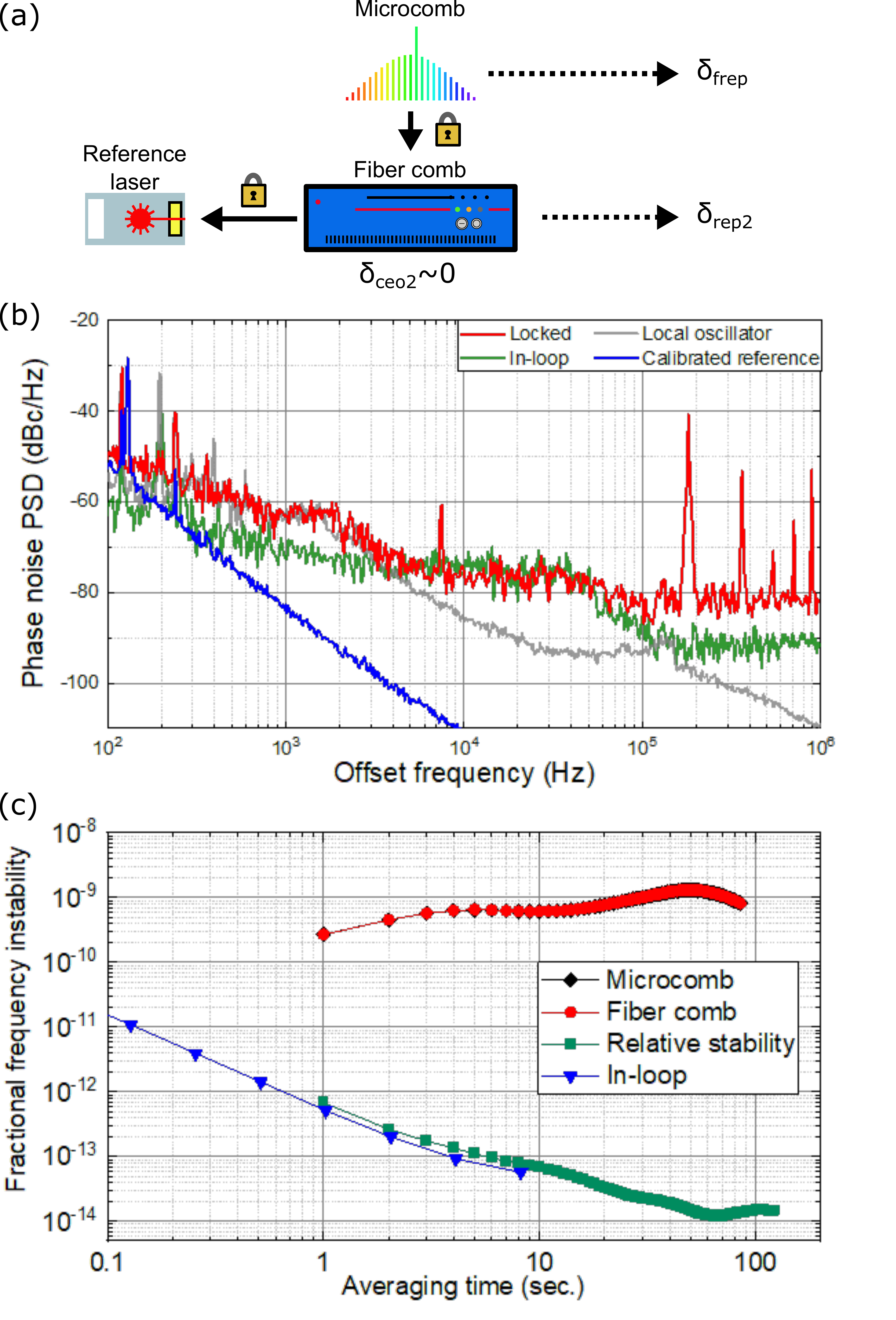}
    \caption{\footnotesize{(\textbf{a}) Conceptional drawing of phase-locking of a microcomb to a stabilized fiber comb. The fiber comb is stabilized to a reference laser. The microcomb is phase-locked to the stabilized fiber comb. (\textbf{b}) Phase noise of a microcomb phase-locked to a stabilized fiber comb. The data below and above 1.7~kHz offset is measured with the SYN and DRO, respectively. (\textbf{c}) Fractional frequency instability in terms of modified Allan deviation for stabilized micro and fiber combs.}}
    \label{fig5}
\end{figure*}

Finally, we stabilize the fiber comb through the optical frequency division (OFD)~\cite{fortier2011generation}, where the CEO noise of the fiber comb is canceled by the f-2f self-referencing scheme and one of its comb lines is phase-locked to a reference laser (ORION laser module, RIO) (Fig.~\ref{fig5}(a)).
We characterize the microcomb's phase noise and frequency stability in the phase-locked condition to the stabilized fiber comb. 
The measured phase noise is plotted in Fig.~\ref{fig5}(b). To maximize the sensitivity, two independent measurement is performed by employing two different local oscillators as the EO comb driver, and the obtained data is stitched. To be specific, we employed the SYN in Fig.~\ref{fig1}(b) for the measurement between 100~Hz and about 1.7~kHz offsets since it has lower phase noise in the range compared to the DRO in Fig.~\ref{fig1}(b), which is used for the measurement at other offset frequencies.
Ideally, the phase noise will be same level as the divided reference laser noise through the OFD (blue line), however, it is close to the level only around the offset frequency of 100~Hz, and limited by the employed local oscillator noise (grey line: below 5~kHz offset), the in-loop noise (green line: about 5~kHz to 60~kHz offset) and the shot noise (above 60~kHz offset). The obtained phase noise at 10~kHz offset is about -75~dBc/Hz. 
The measured frequency instability is shown in Fig.~\ref{fig5}(c). The fractional frequency instability plots of the microcomb (instability of $\delta _\mathrm{f_{rep}}$) and the fiber comb (instability of $\delta _\mathrm{{rep2}}$) are overlapped well thanks to the phase-locked loop. The frequency instability at 1~second is $2.7\times 10^{-10}$. To evaluate how faithfully the microcomb follows the fiber comb, we derive relative frequency instability between the two combs (green line). Although it shows low instability level of $5.2\times 10^{-13}$ at 1~second averaging time, the level is limited by the frequency instability of the in-loop signal (blue line). This gives current limitation of reachable frequency instability level, and further reduction of intrinsic noise of a microcomb (e.g., by resonator dispersion engineering~\cite{stone2020harnessing}) will be required to improve this. Implementation of an alternative actuator for the feedback control with a larger dynamic range is also helpful because that of the actuator employed in this study (the pump power control through the SSBM) is rather small. 

\section{Conclusion}
We experimentally evaluate impacts of pump-resonance detuning and resonator temperature on properties of a microcomb. We observe that resonator temperature can be an important parameter not only for minimizing phase noise but for finding a quiet point within an operational range of pump-resonance detuning. We confirm that frequency instability is minimized at a quiet point as well as phase noise. The obtained lowest fractional frequency instability of $1.5\times 10^{-9}$ is 44 times lower than that of a free-running condition. Effects of noise suppression through the use of a quiet point is examined in a feedback loop for phase-locking a microcomb to a fiber comb. Visible differences in the locking performance are observed between two conditions with different initial noise levels, and the advantage of the method is clearly presented. By stabilizing the fiber comb to a reference laser, we obtain phase noise and frequency stability as low as -75~dBc/Hz at 10~kHz offset and $2.7\times 10^{-10}$ at 1~second, respectively. The obtained relative frequency stability between the microcomb and the fiber comb of $5.2\times 10^{-13}$ at 1~second suggests a lower limit of frequency instability achieved in the current configuration. Engineering of resonator dispersion will be an effective way to lower this level. The empirical investigation demonstrated here shows a helpful strategy to find a quiet condition to stabilize a microcomb.

\bibliography{osa_c1}

\end{document}